\documentclass[a4paper,11pt]{article}
\pdfoutput=1
\usepackage{jheppub}

\usepackage{amssymb}
\usepackage{bm,bbm,amsmath}

\usepackage{graphicx}
\usepackage[export]{adjustbox}

\usepackage{layouts}
\usepackage{xparse}

\usepackage{hyperref}
\usepackage{hypernat}

\usepackage[caption=false,labelfont=normalsize,labelformat=empty,textfont=normalsize,justification=centering]{subfig}

\renewcommand{\Im}{\mathrm{Im}}

\newcommand{\iu}{\mathrm{i}}
\newcommand{\eq}{\mathrm{eq}}
\newcommand{\D}{\mathrm{d}}

\title{\boldmath Asymmetric dark matter from semi-annihilation: unitarity constraints and long-lived final states}

\author[a,2]{Tom\'a\v s Bla\v zek}
\author[a,1]{Peter Mat\'ak\note{Corresponding author.}}
\author[a,3]{Viktor Zaujec}

\affiliation[a]{Department of Theoretical Physics, Comenius University,\\ Mlynsk\'a dolina, 84248 Bratislava, Slovak Republic}

\emailAdd{tomas.blazek@fmph.uniba.sk}
\emailAdd{peter.matak@fmph.uniba.sk}
\emailAdd{viktor.zaujec@fmph.uniba.sk}

\abstract{
This work presents an asymmetric dark matter model with relic density determined by the freeze-out of asymmetric semi-annihilations into long-lived particles slowly decaying into the Standard Model states. We carefully consider the $CPT$ symmetry and unitarity constraints to the asymmetries entering the Boltzmann equation. The main idea of the paper is to point out a critical inconsistency in the previous literature, where these constraints are violated. We present a systematic approach to avoid the inconsistency.
}


\begin{document}

\maketitle
\flushbottom

\section{Introduction.}

According to Sakharov's conditions \cite{Sakharov:1967dj}, $CP$-violating processes are necessary for producing an asymmetry from the initially symmetric universe. Although the most straightforward approaches rely on asymmetric heavy-particle decays \cite{Blennow:2010qp, Haba:2010bm, Davoudiasl:2010am, Gu:2010ft, Falkowski:2011xh, Arina:2011cu, Feng:2013wn, Borah:2024wos}, asymmetries from two-particle annihilations have also been studied in the literature \cite{Farrar:2004qy, Baldes:2014gca, Bell:2014xta, Ghosh:2021qbo}. The present work focuses on semi-annihilations \cite{DEramo:2010keq, DEramo:2012fou, Cai:2015zza, Cai:2016hne, Kamada:2017gfc, Cai:2019pcj, Bandyopadhyay:2022tsf, Guo:2023kqt, Benincasa:2023vyp, Beauchesne:2024vbo}, in which two dark matter particles scatter into a dark matter antiparticle and a particle from the visible sector. We demonstrate that if these visible sector particles are long-lived, the dark sector can become almost totally asymmetric.

In general, the effect of the dark matter asymmetry can be two-fold. A small asymmetry can be transferred to the visible sector via portal interactions, which can lead to successful baryogenesis \cite{Kuzmin:1996he, Kitano:2004sv, Kaplan:2009ag, Petraki:2013wwa, Zurek:2013wia}. Another related aspect is the interplay of asymmetry and dark-matter relic density. Although the symmetric density component can be removed through its annihilations to the Standard Model, even a tiny difference in the densities of dark matter particles and antiparticles can affect the relic abundance \cite{Iminniyaz:2011yp}. We focus on the latter, postponing the possible connection to the visible-sector matter-antimatter asymmetry for future work. Our goal is not to present an entirely accurate extension of the Standard Model. Instead, we study the viability of the asymmetric semi-annihilations as a mechanism.

As the underlying re\-a\-cti\-ons contributing to the asymmetry source term, semi-anni\-hi\-la\-tions were previously considered in Refs.~\cite{Ghosh:2020lma, Ghosh:2021ivn}. Unfortunately, we must disagree with the results of Ref.~\cite{Ghosh:2020lma}, as they violate the $CPT$ and unitarity constraints \cite{Kolb:1979qa, Dolgov:1979mz, Hook:2011tk, Bhattacharya:2011sy, Blazek:2021olf}. For the particular model in Ref.~\cite{Ghosh:2020lma}, the seemingly nonvanishing asymmetry results from omitting a leading-order diagram, which, if included, would interfere with higher orders and completely erase the asymmetry.

The paper is structured as follows. In Section \ref{sec2}, we discuss the $CPT$ symmetry and unitarity constraints, explaining why the previous approach to asymmetric semi-annihilations fails. We explain how $S$-matrix unitarity can be used to recognize vanishing asymmetry at the diagrammatic level and how to prevent such failure. We also show how asymmetries can arise at higher orders and present a systematic procedure to identify all relevant terms. Section \ref{sec3} is devoted to modifying the previous model, considering more than one visible-sector particle species produced in semi-annihilations. We summarize in Section \ref{sec4}.

\section{Unitarity constraints for asymmetric semi-annihilations\label{sec2}}

\begin{figure}
\centering\includegraphics[scale=1]{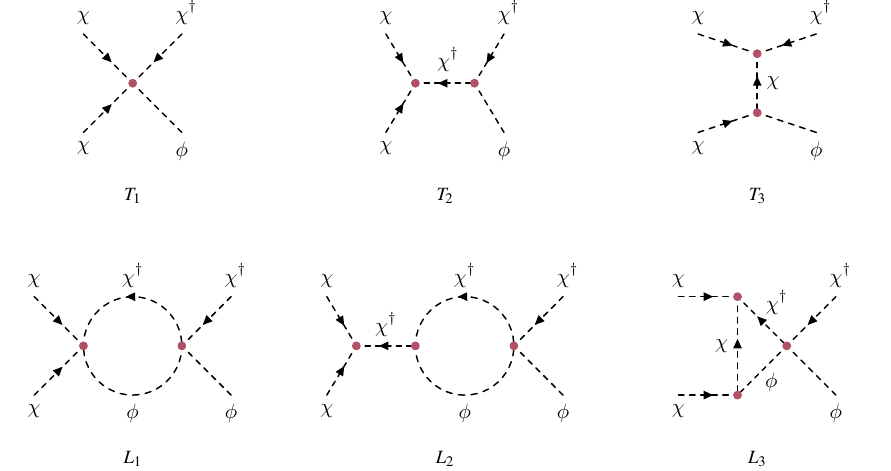}
\caption{Tree and loop diagrams for the dark matter semi-annihilation according to model in Eq.~\eqref{lag1}. Note that our numbering differs from Ref.~\cite{Ghosh:2020lma}, where the diagram $T_3$ was not included.}
\label{fig1}
\end{figure}

Within a quantum field theory, two conditions must be met to violate the $CP$ symmetry in particle interactions. First, irreducible complex phases in coupling must be present in the Lagrangian of the model, and second, imaginary kinematics must occur in the loop integrals contributing to the amplitude of the considered process. But even with these requirements fulfilled, the asymmetries may still vanish due to the $S$-matrix unitarity and the $CPT$ symmetry. Taking the diagonal elements of $S^\dagger\! S=SS^\dagger$ and separating the interacting part as $\iu T=S-\mathbbm{1}$, we obtain \cite{Kolb:1979qa, Dolgov:1979mz}
\begin{align}\label{unit1}
\sum_f\Delta\vert T_{fi}\vert^2=\sum_i\Delta\vert T_{fi}\vert^2=0\quad\text{for}\quad\Delta\vert T_{fi}\vert^2=
\vert T_{fi}\vert^2-\vert T_{if}\vert^2.
\end{align}
The asymmetries thus never come alone. Instead, for a fixed initial state, there must always be at least two reactions of mutually opposite asymmetries cancelling each other. 

The authors in Ref.~\cite{Ghosh:2020lma} claim that dark-matter semi-annihilation is the only asymmetric reaction in their scenario. However, in consequence of Eq.~\eqref{unit1}, such an asymmetry must vanish identically, and indeed it does. To demonstrate that, we consider the model introduced therein, based on the interactions of a complex scalar $\chi$ representing the asymmetric dark matter candidate and a real scalar $\phi$ decaying into or mixing with the Standard Model sector. The theory was illustrated by an example of a $\mathbb{Z}(3)$-symmetric Lagrangian density
\begin{align}\label{lag1}
\mathcal{L}\supset\frac{\mu}{3!}\chi^3+\frac{\lambda}{3!}\phi\chi^3+\text{H.c.}
+\frac{\lambda_1}{4}\vert\chi\vert^4+\frac{\lambda_2}{2}\phi^2\vert\chi\vert^2
+\mu_1\phi\vert\chi\vert^2+\frac{\mu_2}{3!}\phi^3+\frac{\lambda_3}{4!}\phi^4.
\end{align}
The couplings $\mu$ and $\lambda$ are generally complex. Only one of them can be made real by redefining the field $\chi$. We may thus hope for contributions to the $\chi\chi\rightarrow \chi^\dagger\phi$ semi-annihilation asymmetry from the interference of the tree- and loop-level amplitudes labelled by $T$ and $L$ as in Fig.~\ref{fig1}. But a closer look unveils that $T^{\vphantom{*}}_iL^*_i$, for $i=1,2,3$, comes with a real combination of couplings and cannot lead to an asymmetry. Furthermore, the contribution of $T^{\vphantom{*}}_iL^*_j$ is always cancelled by $T^{\vphantom{*}}_jL^*_i$. The diagrams in Fig.~\ref{fig1} thus cannot lead to an asymmetry in semi-annihilations. A nonzero result can only be obtained by mistake, which can easily happen, as the usual asymmetry calculations rely on a separate treatment of imaginary couplings and loop kinematics. The interfering amplitudes are split unnaturally into these two parts, making it difficult to track the asymmetry cancellations required by Eq.~\eqref{unit1}. 

\subsection{Unitarity constraints from holomorphic cutting rules\label{sec3_1}}

To simplify the calculations and prevent mistakes, we can rewrite the $S$-matrix unitarity condition as $\iu T^\dagger=\iu T - \iu T^\dagger \iu T$. Its iterative application to $\vert T_{fi}^{\vphantom{\dagger}}\vert^2=-\iu T^\dagger_{if}\iu T^{\vphantom{\dagger}}_{fi}$ yields a general formula \cite{Blazek:2021olf}
\begin{align}\label{unit2}
\Delta\vert T^{\vphantom{\dagger}}_{fi}\vert^2=
\vert T^{\vphantom{\dagger}}_{fi}\vert^2-\vert T^{\vphantom{\dagger}}_{if}\vert^2
=&\sum_{n}\Big(\iu T^{\vphantom{\dagger}}_{\vphantom{f}in}\iu T^{\vphantom{\dagger}}_{nf}\iu T^{\vphantom{\dagger}}_{fi}
-\iu T^{\vphantom{\dagger}}_{\vphantom{f}if}\iu T^{\vphantom{\dagger}}_{fn}\iu T^{\vphantom{\dagger}}_{ni}\Big)\\
&-\sum_{n,k}\Big(\iu T^{\vphantom{\dagger}}_{\vphantom{f}in}\iu T^{\vphantom{\dagger}}_{\vphantom{f}nk}\iu T^{\vphantom{\dagger}}_{kf}\iu T^{\vphantom{\dagger}}_{fi}
-\iu T^{\vphantom{\dagger}}_{if}\iu T^{\vphantom{\dagger}}_{fk}\iu T^{\vphantom{\dagger}}_{kn\vphantom{f}}\iu T^{\vphantom{\dagger}}_{\vphantom{f}ni}\Big)\nonumber\\
&+\ldots\nonumber
\end{align}
The asymmetries are thus written as a product of at least three diagrams or as a forward-scattering diagram with two or more holomorphic cuts \cite{Coster:1970jy, Bourjaily:2020wvq, Hannesdottir:2022bmo}. For the lowest orders, one may alternatively use the approach of Ref.~\cite{Roulet:1997xa}, representing the asymmetries by cyclic diagrams. The contribution analogous to the interference of $T^{\vphantom{*}}_1$ and $L^*_3$ diagrams in Fig.~\ref{fig1} ($T^{\vphantom{*}}_1$ and $L^*_1$ according to the notation of Ref.~\cite{Ghosh:2020lma}) then corresponds to
\begin{align}
\includegraphics[valign=c,scale=1]{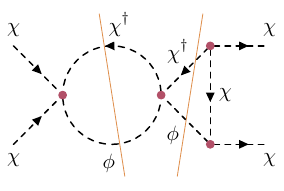}\hskip2mm-\hskip2mm\includegraphics[valign=c,scale=1]{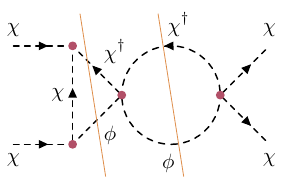}
\end{align}
which is cancelled by 
\begin{align}
\includegraphics[valign=c,scale=1]{math1b.pdf}\hskip2mm-\hskip2mm\includegraphics[valign=c,scale=1]{math1a.pdf}
\end{align}
obtained from $T^{\vphantom{*}}_3$ and $L^*_1$. It is also easy to see that for all diagrams in Fig.~\ref{fig1}, the situation is the same, and, as already argued, no asymmetry in semi-annihilations occurs. 

\subsection{Higher-order {\it CP} asymmetries and vacuum diagrams\label{sec2_2}}

Despite the absence of asymmetry found in the previous subsection, we may still use the irreducible phase of Eq.~\eqref{lag1} to identify reactions with $CP$ asymmetric rates. As seen in Eq.~\eqref{unit2}, asymmetries come from forward-scattering diagrams cut into multiple pieces. Therefore, as a starting point, we may take a certain set of vertices and draw all possible vacuum diagrams out of them. Following the procedure of Ref.~\cite{Blazek:2021olf}, we may cut the vacuum diagrams in all possible ways, making forward-scattering diagrams that are further cut to obtain transition probabilities via the optical theorem. Once entering the right-hand side of the Boltzmann equation, they include not only the rates attributed to the usual connected reaction topologies---decays, scatterings, or semi-annihilations---but also may approximate the effects of quantum statistics and thermal masses \cite{Blazek:2021zoj, Blazek:2021gmw, Blazek:2022azr}. 

Vacuum diagrams have also been shown to represent rephasing invariants \cite{Botella:2004ks}, that is, combinations of couplings invariant under the field redefinitions. Such combinations are irreducibly complex if and only if reversing the arrows on the internal lines of the respective vacuum diagram produces a new vacuum diagram \cite{Roulet:1997xa}. 

\begin{figure}
\subfloat{\label{fig2a}}
\subfloat{\label{fig2b}}
\subfloat{\label{fig2c}}
\subfloat{\label{fig2d}}
\subfloat{\label{fig2e}}
\centering\includegraphics[width=\linewidth]{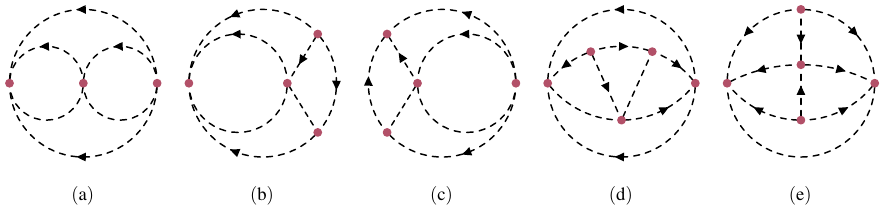}
\caption{Example vacuum diagrams allowed by the Lagrangian density in Eq.~\eqref{lag1}. Lines with or without arrow correspond to $\chi$ or $\phi$ propagators, respectively. Here, the asymmetries can only arrise from the diagrams in Figs.~\ref{fig2d} and \ref{fig2e}, which are not invariant under the arrow reversal and contribute to multiple processes.}
\label{fig2}
\end{figure}

For the model in Eq.~\eqref{lag1}, the only complex phase comes from $\lambda\mu^*$. An asymmetry can only originate from diagrams that contain some power of these two couplings. In Fig.~\ref{fig2a}, we can see the vacuum diagram made of $T^{\vphantom{*}}_1$ and $L^*_1$ proportional to $\vert\lambda\vert^2\lambda_2$ that cannot lead to asymmetry. Reversing the arrows on the $\chi$ internal lines yields the same diagram. The same operation exchanges the diagrams in Figs.~\ref{fig2b} and \ref{fig2c}. However, both contribute to the asymmetry of the same reaction $\chi\chi\rightarrow\chi^\dagger\phi$ and therefore cancel each other out, as we have seen in the previous section. Finally, the diagrams in Figs.~\ref{fig2d} and \ref{fig2e} come with $\lambda\mu^* \lambda_1\lambda_2\mu_1$ and $\lambda^2\mu^{*2}\lambda_1$, respectively. Both products are complex, and the arrow reversal leads to new diagrams. As an example, we may cut the diagram in Fig.~\ref{fig2e} and obtain
\begin{align}\label{math3ab}
\includegraphics[valign=c,scale=1]{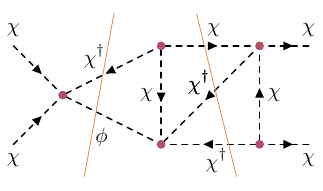}\hskip2mm-\hskip2mm\includegraphics[valign=c,scale=1]{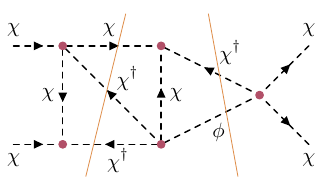}
\end{align}
leading to an asymmetry in $\chi\chi\rightarrow\chi^\dagger\phi$ reaction. 
From the second term in Eq.~\eqref{math3ab}, we can see that the latter is canceled by the contribution of the opposite sign to $\chi\chi\rightarrow \chi^\dagger\chi^\dagger\chi$. We may also permute the tree-level diagram pieces in each term of Eq.~\eqref{math3ab} and obtain the asymmetry in the rates of $\chi^\dagger\phi\rightarrow\chi^\dagger\chi^\dagger\chi$ and $\chi^\dagger\chi^\dagger \chi\rightarrow\chi\chi$. The former preserves the $\chi$ particle number and does not contribute to the asymmetry source term, while the latter appears in Ref.~\cite{Ghosh:2021ivn} along with other reactions with $\chi\chi$ in the initial or final state.

The same procedure can be applied to a diagram with one more line cut for the initial state, such as
\begin{align}\label{math3cd}
\includegraphics[valign=c,scale=1]{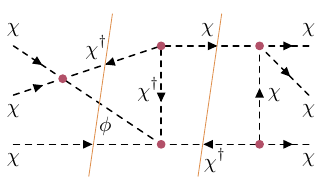}\hskip2mm-\hskip2mm\includegraphics[valign=c,scale=1]{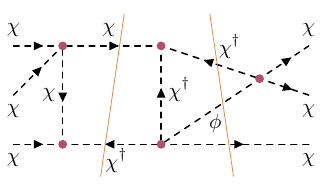}  
\end{align}
contributing to asymmetries in $3\chi\rightarrow\chi^\dagger\phi\chi$, $\chi^\dagger\phi\chi\rightarrow \chi^\dagger\chi$, and $\chi^\dagger\chi\rightarrow 3\chi$.
These come with no more phase-space or Boltzmann suppression than the asymmetries obtained from Eq.~\eqref{math3ab}. The asymmetry contributions from Eq.~\eqref{math3cd} are therefore equally important as the semi-annihilation asymmetry in Eq.~\eqref{math3ab}, and should be included in the Boltzmann equation. 

A patient reader may surely draw tens of complex vacuum diagrams made of five or more vertices within the model in Eq.~\eqref{lag1} that lead to possibly hundreds of independent asymmetry contributions. We instead suggest avoiding this tedious work and modifying the model to allow for asymmetric semi-annihilations at lower orders.

\section{Asymmetric semi-annihilations into long-lived particles\label{sec3}}

The asymmetry in a certain process can be obtained if there are at least two different final states for a fixed initial state. For semi-annihilations, this can be achieved by including multiple $\chi$ or $\phi$ fields. We consider the second option, first, in a purely scalar theory, similar to Eq.~\eqref{lag1}. Then, we explore the properties of an analogous fermionic model with effective four-fermion interactions.

\subsection{Scalar dark matter with asymmetric semi-annihilations}
Let us consider the Lagrangian density
\begin{align}\label{lag2}
\mathcal{L}\supset-\frac{\lambda_1}{6}\chi^3\phi_1-\frac{\lambda_2}{6}\chi^3\phi_2
+\text{H.c.}-\lambda_{12}\vert\chi\vert^2\phi_1\phi_2-\frac{\lambda_3}{2}\vert\chi\vert^2\phi^2_3
\end{align}
where $\chi$ is again a dark sector particle of mass $m$ charged under $\mathbb{Z}(3)$ and $\phi_i$ ($i=1,2,3$) are real scalars. The masses and widths of $\phi_1$ and $\phi_2$ are denoted $m_1$, $m_2$, and $\Gamma_1$, $\Gamma_2$, respectively. We do not specify their decay channels and only assume a slow decay to the Standard Model. The last term in Eq.~\eqref{lag2} stands for dark-matter coupling to the heat bath represented by a massless $\phi_3$, leading to fast dark-matter annihilations removing the symmetric density component. 

We can adjust the phase of the field $\chi$ to make $\lambda_1$ real while $\lambda_2$ remains complex, or vice versa. The invariant combination is $\lambda^*_1\lambda^{\vphantom{*}}_2$ and the asymmetry will be proportional to its imaginary part. For semi-annihilations, let us consider
\begin{align}\label{asym}
\includegraphics[valign=c,scale=1]{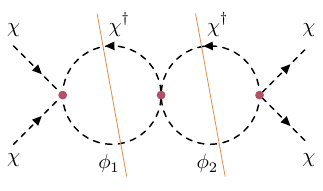}\hskip2mm-\hskip2mm\includegraphics[valign=c,scale=1]{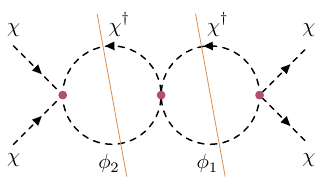}
\end{align}
which, by Eq.~\eqref{unit2}, corresponds to the asymmetry of $\chi\chi\rightarrow\chi^\dagger\phi_1$ reaction. A similar contribution to $\chi\chi\rightarrow\chi^\dagger\phi_2$ will come from the same diagrams in reverse order. 

Thermally averaged cross sections can be computed following the approach of Ref.~\cite{Gondolo:1990dk}. The semi-annihilation asymmetry then yields
\begin{align}\label{DeltaCS}
\Delta\langle\sigma v\rangle_{\chi\chi\rightarrow\chi^\dagger\phi_1}=
\frac{\Im[\lambda^*_1\lambda^{\vphantom{*}}_2]\lambda^{\vphantom{*}}_{12}}{64\pi^2}
\int^\infty_{s_{\min}}\D s\, \sqrt{s-4m^2}
\frac{P_{\chi\phi_1}(s)P_{\chi\phi_2}(s)}{2sm^4}
\frac{K_1(\sqrt{s}/T)}{TK^2_2(m/T)}
\end{align}
where $s_{\min}=\max[4m^2,(m+m_1)^2,(m+m_2)^2]$, $K_1$ and $K_2$ are modified Bessel functions of the second kind, and
\begin{align}
P_{\chi\phi_i}(s)=\frac{1}{2\sqrt{s}}\big[s-(m+m_i)^2\big]^{1/2}\big[s-(m-m_i)^2\big]^{1/2}.
\end{align}
The integral in Eq.~\eqref{DeltaCS} is symmetric under exchange $\phi_1\leftrightarrow\phi_2$. Therefore, it is now explicit that the asymmetry of $\chi\chi\rightarrow\chi^\dagger\phi_2$ only differs by the imaginary part of the couplings and, compared to Eq.~\eqref{DeltaCS}, is of the opposite sign.

To write down the Boltzmann equations, it is convenient to introduce $Y=(n_{\chi\vphantom{\Bar{\chi}}}+ n_{\Bar{\chi}})/s$, $\Delta=(n_{\chi\vphantom{\Bar{\chi}}}-n_{\Bar{\chi}})/s$, and $Y_{\phi_i}=n_{\phi_i}/s$, where $n$s are the number densities and $s$ is the entropy density of the thermal medium. The evolution of the asymmetry comes from two competing parts---source and washout terms---equal to
\begin{align}
\bigg(\frac{\D\Delta}{\D x}\bigg)_{\text{source}}=
&-\frac{3}{8}\frac{s}{Hx}YY^\eq_0\bigg(\frac{Y_{\phi_1}}{Y^\eq_{\phi_1}}-\frac{Y_{\phi_2}}{Y^\eq_{\phi_2}}\bigg)
\Delta\langle\sigma v\rangle_{\chi\chi\rightarrow\chi^\dagger\phi_1}\label{source}\\
\bigg(\frac{\D\Delta}{\D x}\bigg)_{\text{washout}}=
&-\frac{3}{2}\frac{s}{Hx}\Delta Y^\eq_0\sum_{i=1,2}
\bigg(\frac{Y}{Y^\eq_0}+\frac{1}{2}\frac{Y_{\phi_i}}{Y^\eq_{\phi_i}}\bigg)\langle\sigma v\rangle_{\chi\chi\rightarrow\chi^\dagger\phi_i}\label{washout}
\end{align}
where $H$ is the Hubble parameter, $x=m/T$, and $Y^\eq_0$ is the equilibrium value of $Y$ at zero chemical potential (see the discussion on page 6 in Ref.~\cite{Ghosh:2020lma}). From Eq.~\eqref{source}, we can observe that the ratios of the $\phi_1$ and $\phi_2$ densities to their equilibrium values must differ. Otherwise, the contributions of the two reactions cancel each other out. This observation disagrees with what has been found in Ref.~\cite{Ghosh:2020lma}, where the asymmetry was present even for the single scalar $\phi$ in thermal equilibrium. As we can see from Eq.~\eqref{washout}, the asymmetry would be quickly washed out in that case.

The Boltzmann equations for $Y$ and $Y_{\phi_i}$ receive contributions from annihilations, semi-annihilations, and decays and can be written as
\begin{align}
\frac{\D Y}{\D x}=\label{Y}
&-\frac{1}{2}\frac{s}{Hx}\bigg[\sum_{i=1,2}\frac{1}{2}\bigg(Y^2+\Delta^2-YY^\eq_0\frac{Y_{\phi_i}}{Y^\eq_{\phi_i}}\bigg) \langle\sigma v\rangle_{\chi\chi\rightarrow\chi^\dagger\phi_i}\\
&+\bigg(Y^2-\Delta^2-Y^{\eq 2}_0 
\frac{Y_{\phi_1}}{Y^\eq_{\phi_1}}\frac{Y_{\phi_2}}{Y^\eq_{\phi_2}}\bigg)
\langle\sigma v\rangle_{\chi\chi^\dagger\rightarrow\phi_1\phi_2}+\bigg(Y^2-\Delta^2 -Y^{\eq 2}_0\bigg)
\langle\sigma v\rangle_{\chi\chi^\dagger\rightarrow\phi_3\phi_3}\bigg]\nonumber\\
\frac{\D Y_{\phi_1}}{\D x}=\label{Yphi1}
&-\frac{\langle\Gamma_1\rangle}{Hx}\bigg(Y^{\vphantom{\eq}}_{\phi_1}-Y^\eq_{\phi_1}\bigg)
+\frac{1}{4}\frac{s}{Hx}\bigg[\bigg(Y^2+\Delta^2-YY^\eq_0\frac{Y_{\phi_1}}{Y^\eq_{\phi_1}}\bigg) \langle\sigma v\rangle_{\chi\chi\rightarrow\chi^\dagger\phi_1}\\
&+\bigg(Y^2-\Delta^2-Y^{\eq 2}_0 
\frac{Y_{\phi_2}}{Y^\eq_{\phi_2}}\frac{Y_{\phi_1}}{Y^\eq_{\phi_1}}\bigg)
\langle\sigma v\rangle_{\chi\chi^\dagger\rightarrow\phi_1\phi_2} -4YY^\eq_{\phi_1}\bigg(\frac{Y_{\phi_1}}{Y^\eq_{\phi_1}}-\frac{Y_{\phi_2}}{Y^\eq_{\phi_2}}\bigg)\langle\sigma v\rangle_{\chi\phi_1\rightarrow\chi\phi_2} \bigg]\nonumber
\end{align}
and similarly for $\phi_2$. Note that we neglect the asymmetry of $\chi\phi_2\rightarrow\chi\phi_1$ in the last term of Eq.~\eqref{Yphi1}. 

\begin{figure}
\subfloat{\label{fig3a}}
\subfloat{\label{fig3b}}
\centering\includegraphics[width=\linewidth]{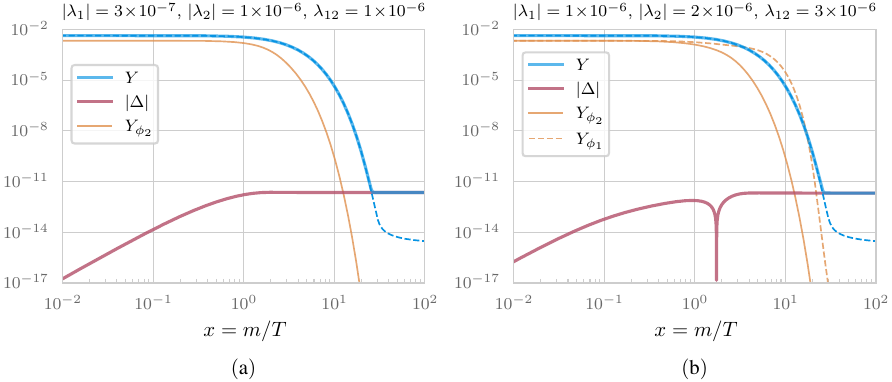}
\caption{The evolution of $Y$, $\vert\Delta\vert$, $Y_{\phi_2}$ , and $Y_{\phi_1}$ following from Eqs.~\eqref{source}-\eqref{Yphi1}. Values of $Y$ computed solely from $\chi\chi^\dagger \rightarrow\phi_3\phi_3$ with $\lambda_3=6.0$ are plotted by the dashed blue lines and serve for comparison to what would be obtained if the effects of asymmetry were neglected. In both cases, $m=200~\mathrm{GeV}$, $m_2=400~\mathrm{GeV}$, $\Gamma_2/m_2=3\times 10^{-16}$, and $\Im[\lambda^{\vphantom{*}}_1\lambda^*_2]/\vert \lambda^*_1\lambda^{\vphantom{*}}_2\vert=0.8$. We put $m_1=0$ with $\phi_1$ in thermal equilibrium (left), or $m_1=260~\mathrm{GeV}$ and $\Gamma_1/m_1=2\times 10^{-17}$ (right).}
\label{fig3}
\end{figure}

Numerical solutions of the Boltzmann equations for two sets of parameter values are shown in Fig.~\ref{fig3}. In both cases, total asymmetry was achieved for the dark-matter relic density corresponding to the observed value. That can only happen for significant deviations from thermal equilibrium for at least one of the $\phi_1$, $\phi_2$ particles. In other words, at least one of them has to be long-lived. 

In Fig.~\ref{fig3a}, we assumed that $\phi_1$ is in thermal equilibrium and $m_1$ is negligible compared to $m=200~\mathrm{GeV}$ and $m_2=400~\mathrm{GeV}$. Furthermore, all $\phi_2$ interactions with $\phi_1$ and $\chi$ had to be suppressed to prevent equilibration. The asymmetry had been slowly produced and had frozen in before the dark-matter density froze out. 

The situation in Fig.~\ref{fig3b} is more complicated. First, asymmetry production was dominated by out-of-equilibrium effects of $\phi_2$. Then, as the temperature decreased, massive $\phi_1$ decayed too slowly to follow the equilibrium distribution and also absorbed the remaining $\phi_2$ through $\chi\phi_2 \rightarrow\chi\phi_1$ conversions. The asymmetry changed sign and froze in at $x\simeq 3$. 

Naively, one might have expected the relic density to depend on two-particle annihilations only, as we used large $\lambda_3$, while other couplings were smaller by more than five orders of magnitude. However, contrary to expectation, the asymmetry produced was large enough to increase the relic density by multiple orders of magnitude.

\subsection{Asymmetric semi-annihilations of fermions}

For scalar particles, large asymmetries were obtained for semi-anni\-hi\-la\-tions through feeble interactions. However, small dimensionless couplings may be considered tuned or unnatural. In this section, we rephrase the smallness of the couplings in terms of a high-energy scale suppressing the contribution of higher-dimensional effective operators within a fermionic model analogous to Eq.~\eqref{lag2}. Let us consider
\begin{align}\label{lag3}
\mathcal{L}\supset
&-\frac{\kappa_1}{6\Lambda^2}(\Bar{\chi}^cP_R\chi)(\Bar{\chi}^cP_L\psi_1)
-\frac{\kappa_2}{6\Lambda^2}(\Bar{\chi}^cP_R\chi)(\Bar{\chi}^cP_L\psi_2)\\
&-\frac{\kappa_{12}}{2\Lambda^2}(\Bar{\chi}^cP_L\psi_1)^\dagger(\Bar{\chi}^cP_L\psi_2)
+\text{H.c.}\nonumber
\end{align}
where $\chi$ is now a Dirac fermion and the dark matter candidate, while $\psi_1$ and $\psi_2$ are Majorana fermions. Note that the projectors must have opposite chiralities in each of the first two terms in Eq.~\eqref{lag3}. Otherwise, the four-fermion operators would vanish even at the classical level, since the components of the classical field $\chi$ are Grassmannian numbers.

\begin{figure}
\subfloat{\label{fig4a}}
\subfloat{\label{fig4b}}
\centering\includegraphics[width=\linewidth]{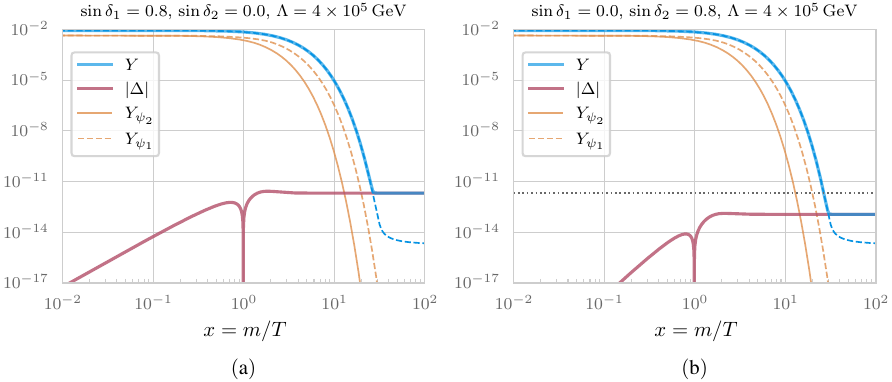}
\caption{Numerical solutions for particle densities within the fermionic model. Up to a complex phase, $\kappa_1$, $\kappa_2$ and $\kappa_{12}$ were set to unity, while $y=3.0$, $m=200~\mathrm{GeV}$, $m_1=260~\mathrm{GeV}$, $m_2=400~\mathrm{GeV}$, $\Gamma_1/m_1=1\times 10^{-15}$ and $\Gamma_2/m_2=4\times 10^{-15}$. In the left panel, the parameters were tuned to reproduce the observed dark matter density, represented by the grey dotted line in the right panel. Similarly to Fig.~\ref{fig3}, the blue dashed line corresponds to density evolution from annihilations only.} 
\label{fig4}
\end{figure}

The semi-annihilation asymmetries come from forward-scattering diagrams similar to those in Eq.~\eqref{asym}. Here the spin sums in squared amplitudes were evaluated using {\texttt{FeynCalc}} \cite{Shtabovenko:2020gxv, Shtabovenko:2016sxi, Mertig:1990an}. Remarkably, Eq.~\eqref{lag3} allows for two different clippings of the Majorana fields in the $\chi\psi_1 \rightarrow\chi\psi_2$ tree-level amplitude, proportional to $\kappa^{\vphantom{*}}_{12}$ and $\kappa^*_{12}$. While in the scalar model, $\lambda_{12}$ was real, here $\kappa^{\vphantom{*}}_{12}$ is complex and leads to two independent phases. The asymmetry of the thermally averaged cross section therefore contains two terms, proportional to $\Im[\kappa^*_1 \kappa^{*}_{12}\kappa^{\vphantom{*}}_2]=\vert\kappa^*_1 \kappa^{*}_{12}\kappa^{\vphantom{*}}_2\vert\sin\delta_1$ and $\Im[\kappa^*_1\kappa^{\vphantom{*}}_{12}\kappa^{\vphantom{*}}_2]=\vert\kappa^*_1\kappa^{\vphantom{*}}_{12}\kappa^{\vphantom{*}}_2\vert\sin\delta_2$. 

The Lagrangian density in Eq.~\eqref{lag3} leads to the same reactions as were obtained from Eq.~\eqref{lag2}. We similarly define $Y$, $\Delta$, $Y_{\psi_1}$, and $Y_{\psi_2}$, and their evolution will follow from the same set of Boltzmann equations as in the previous section. Furthermore, we also include the Yukawa interaction of the $\chi$ field with the real scalar $\phi_3$, represented by $\mathcal{L}\supset-y\bar{\chi}\chi\phi_3$. This will not be suppressed by the high-energy scale $\Lambda$, allowing for a strong annihilation of the symmetric density component.

The numerical solutions of the Boltzmann equations are shown in Fig.~\ref{fig4}. We used the same particle masses as in Fig.~\ref{fig3b}, and in fact, the solutions look similar. The two panels differ only by the role of the two complex phases entering the asymmetry source term. Interestingly, the effect of $\delta_2$ is somewhat limited, and the resulting asymmetry in Fig.~\ref{fig4b} is much lower than in Fig.~\ref{fig4a}, in which the relic density corresponds to the observed value. The results were obtained with $\vert\kappa_1\vert=\vert\kappa_2\vert=\vert\kappa_{12}\vert=1$. Despite suppression by powers of $\Lambda=4\times 10^5~\mathrm{GeV}$, the asymmetry from semi-annihilations increased the relic density by up to three orders of magnitude.

\section{Summary\label{sec4}}

In this paper, we discuss in detail how unitarity and $CPT$ symmetry restrict the properties of the asymmetric dark-matter model based on semi-annihilations. It has been demonstrated that, in agreement with Refs.~\cite{Kolb:1979qa, Dolgov:1979mz} and in contrast to the previous attempt in Ref.~\cite{Ghosh:2020lma}, at least two different final-state processes are needed for a non-zero asymmetry. Therefore, we have introduced two real scalars or Majorana fermions to which the dark matter semi-annihilated. If these particles were long-lived and feebly interacting, with small dimensionless couplings or via higher-dimensional effective operators suppressed by a high-energy scale, the total asymmetry in the dark sector would be achieved.

\section*{Acknowledgements}
We thank A. Ghosh, D. Ghosh, and S. Mukhopadhyay for valuable feedback and discussion. T. Bla\v{z}ek, P. Mat\'{a}k, and V. Zaujec were supported by the Slovak Grant Agency VEGA, project No. 1/0719/23, and Slovak Education Ministry Contract No. 0466/2022. V. Zaujec received funding from Comenius University in Bratislava, grant No. UK/3122/2024.

\bibliographystyle{JHEP.bst}
\bibliography{CLANOK.bib}

\providecommand{\noopsort}[1]{}\providecommand{\singleletter}[1]{#1}%

\providecommand{\href}[2]{#2}\begingroup\raggedright\begin{thebibliography}{10}

\bibitem{Sakharov:1967dj}
A.D.~Sakharov, \emph{{Violation of CP invariance, C asymmetry, and baryon
  asymmetry of the universe}},
  \href{https://doi.org/10.1070/PU1991v034n05ABEH002497}{\emph{Pisma Zh. Eksp.
  Teor. Fiz.} {\bfseries 5} (1967) 32}.

\bibitem{Blennow:2010qp}
M.~Blennow, B.~Dasgupta, E.~Fernandez-Martinez and N.~Rius, \emph{Aidnogenesis
  via leptogenesis and dark sphalerons},
  \href{https://doi.org/10.1007/JHEP03(2011)014}{\emph{Journal of High Energy
  Physics} {\bfseries 2011} (2011) 14}
  [\href{https://arxiv.org/abs/1009.3159}{{\ttfamily 1009.3159}}].

\bibitem{Haba:2010bm}
N.~Haba and S.~Matsumoto, \emph{{Baryogenesis from Dark Sector}},
  \href{https://doi.org/10.1143/PTP.125.1311}{\emph{Progress of Theoretical
  Physics} {\bfseries 125} (2011) 1311}
  [\href{https://arxiv.org/abs/1008.2487}{{\ttfamily 1008.2487}}].

\bibitem{Davoudiasl:2010am}
H.~Davoudiasl, D.E.~Morrissey, K.~Sigurdson and S.~Tulin, \emph{Unified origin
  for baryonic visible matter and antibaryonic dark matter},
  \href{https://doi.org/10.1103/PhysRevLett.105.211304}{\emph{Phys. Rev. Lett.}
  {\bfseries 105} (2010) 211304}
  [\href{https://arxiv.org/abs/1008.2399}{{\ttfamily 1008.2399}}].

\bibitem{Gu:2010ft}
P.-H.~Gu, M.~Lindner, U.~Sarkar and X.~Zhang, \emph{Weakly interacting dark
  matter and baryogenesis},
  \href{https://doi.org/10.1103/PhysRevD.83.055008}{\emph{Phys. Rev. D}
  {\bfseries 83} (2011) 055008}
  [\href{https://arxiv.org/abs/1009.2690}{{\ttfamily 1009.2690}}].

\bibitem{Falkowski:2011xh}
A.~Falkowski, J.T.~Ruderman and T.~Volansky, \emph{Asymmetric dark matter from
  leptogenesis}, \href{https://doi.org/10.1007/JHEP05(2011)106}{\emph{Journal
  of High Energy Physics} {\bfseries 2011} (2011) 106}
  [\href{https://arxiv.org/abs/1101.4936}{{\ttfamily 1101.4936}}].

\bibitem{Arina:2011cu}
C.~Arina and N.~Sahu, \emph{Asymmetric inelastic inert doublet dark matter from
  triplet scalar leptogenesis},
  \href{https://doi.org/https://doi.org/10.1016/j.nuclphysb.2011.09.014}{\emph{Nuclear
  Physics B} {\bfseries 854} (2012) 666}
  [\href{https://arxiv.org/abs/1108.3967}{{\ttfamily 1108.3967}}].

\bibitem{Feng:2013wn}
W.-Z.~Feng, A.~Mazumdar and P.~Nath, \emph{Baryogenesis from dark matter},
  \href{https://doi.org/10.1103/PhysRevD.88.036014}{\emph{Phys. Rev. D}
  {\bfseries 88} (2013) 036014}
  [\href{https://arxiv.org/abs/1302.0012}{{\ttfamily 1302.0012}}].

\bibitem{Borah:2024wos}
D.~Borah, S.~Mahapatra, P.K.~Paul, N.~Sahu and P.~Shukla, \emph{Asymmetric
  self-interacting dark matter with a canonical seesaw model},
  \href{https://doi.org/10.1103/PhysRevD.110.035033}{\emph{Phys. Rev. D}
  {\bfseries 110} (2024) 035033}
  [\href{https://arxiv.org/abs/2404.14912}{{\ttfamily 2404.14912}}].

\bibitem{Farrar:2004qy}
G.R.~Farrar and G.~Zaharijas, \emph{Dark matter and the baryon asymmetry of the
  universe}, \href{https://doi.org/10.1103/PhysRevLett.96.041302}{\emph{Phys.
  Rev. Lett.} {\bfseries 96} (2006) 041302}.

\bibitem{Baldes:2014gca}
I.~Baldes, N.F.~Bell, K.~Petraki and R.R.~Volkas, \emph{Particle-antiparticle
  asymmetries from annihilations},
  \href{https://doi.org/10.1103/PhysRevLett.113.181601}{\emph{Phys. Rev. Lett.}
  {\bfseries 113} (2014) 181601}
  [\href{https://arxiv.org/abs/1407.4566}{{\ttfamily 1407.4566}}].

\bibitem{Bell:2014xta}
N.F.~Bell, S.~Horiuchi and I.M.~Shoemaker, \emph{Annihilating asymmetric dark
  matter}, \href{https://doi.org/10.1103/PhysRevD.91.023505}{\emph{Phys. Rev.
  D} {\bfseries 91} (2015) 023505}
  [\href{https://arxiv.org/abs/1408.5142}{{\ttfamily 1408.5142}}].

\bibitem{Ghosh:2021qbo}
A.~Ghosh, D.~Ghosh and S.~Mukhopadhyay, \emph{Cosmology of complex scalar dark
  matter: Interplay of self-scattering and annihilation},
  \href{https://doi.org/10.1103/PhysRevD.104.123543}{\emph{Phys. Rev. D}
  {\bfseries 104} (2021) 123543}
  [\href{https://arxiv.org/abs/2103.14009}{{\ttfamily 2103.14009}}].

\bibitem{DEramo:2010keq}
F.~D'Eramo and J.~Thaler, \emph{Semi-annihilation of dark matter},
  \href{https://doi.org/10.1007/JHEP06(2010)109}{\emph{Journal of High Energy
  Physics} {\bfseries 2010} (2010) 109}
  [\href{https://arxiv.org/abs/1003.5912}{{\ttfamily 1003.5912}}].

\bibitem{DEramo:2012fou}
F.~D'Eramo, M.~McCullough and J.~Thaler, \emph{Multiple gamma lines from
  semi-annihilation},  \href{https://arxiv.org/abs/1210.7817}{{\ttfamily
  1210.7817}}.

\bibitem{Cai:2015zza}
Y.~Cai and A.~Spray, \emph{Fermionic semi-annihilating dark matter},
  \href{https://doi.org/10.1007/JHEP01(2016)087}{\emph{Journal of High Energy
  Physics} {\bfseries 2016} (2016) 87}
  [\href{https://arxiv.org/abs/1509.08481}{{\ttfamily 1509.08481}}].

\bibitem{Cai:2016hne}
Y.~Cai and A.~Spray, \emph{A systematic effective operator analysis of
  semi-annihilating dark matter},
  \href{https://doi.org/10.1007/JHEP02(2017)120}{\emph{Journal of High Energy
  Physics} {\bfseries 2017} (2017) 120}
  [\href{https://arxiv.org/abs/1611.09360}{{\ttfamily 1611.09360}}].

\bibitem{Kamada:2017gfc}
A.~Kamada, H.J.~Kim, H.~Kim and T.~Sekiguchi, \emph{Self-heating dark matter
  via semiannihilation},
  \href{https://doi.org/10.1103/PhysRevLett.120.131802}{\emph{Phys. Rev. Lett.}
  {\bfseries 120} (2018) 131802}
  [\href{https://arxiv.org/abs/1707.09238}{{\ttfamily 1707.09238}}].

\bibitem{Cai:2019pcj}
H.~Cai, \emph{Radiative neutrino model with semiannihilating dark matter},
  \href{https://doi.org/10.1103/PhysRevD.101.035006}{\emph{Phys. Rev. D}
  {\bfseries 101} (2020) 035006}
  [\href{https://arxiv.org/abs/1907.11595}{{\ttfamily 1907.11595}}].

\bibitem{Bandyopadhyay:2022tsf}
P.~Bandyopadhyay, D.~Choudhury and D.~Sachdeva, \emph{Semiannihilation of
  fermionic dark matter},
  \href{https://doi.org/10.1103/PhysRevD.107.015020}{\emph{Phys. Rev. D}
  {\bfseries 107} (2023) 015020}
  [\href{https://arxiv.org/abs/2206.05811}{{\ttfamily 2206.05811}}].

\bibitem{Guo:2023kqt}
J.~Guo, L.~Wu and B.~Zhu, \emph{Mev gamma-ray constraints for light dark matter
  from semi-annihilation},
  \href{https://doi.org/https://doi.org/10.1016/j.physletb.2023.137853}{\emph{Physics
  Letters B} {\bfseries 840} (2023) 137853}
  [\href{https://arxiv.org/abs/2302.06159}{{\ttfamily 2302.06159}}].

\bibitem{Benincasa:2023vyp}
N.~Benincasa, A.~Hryczuk, K.~Kannike and M.~Laletin, \emph{Phase transitions
  and gravitational waves in a model of {$\mathbb{Z}(3)$} scalar dark matter},
  \href{https://doi.org/10.1007/JHEP02(2024)207}{\emph{Journal of High Energy
  Physics} {\bfseries 2024} (2024) 207}
  [\href{https://arxiv.org/abs/2312.04627}{{\ttfamily 2312.04627}}].

\bibitem{Beauchesne:2024vbo}
H.~Beauchesne and C.-W.~Chiang, \emph{Dark matter semi-annihilation for inert
  scalar multiplets},
  \href{https://doi.org/10.1007/JHEP06(2024)164}{\emph{Journal of High Energy
  Physics} {\bfseries 2024} (2024) 164}
  [\href{https://arxiv.org/abs/2403.01729}{{\ttfamily 2403.01729}}].

\bibitem{Kuzmin:1996he}
V.A.~Kuzmin, \emph{{A Simultaneous solution to baryogenesis and dark matter
  problems}}, \href{https://doi.org/10.1134/1.953070}{\emph{Phys. Part. Nucl.}
  {\bfseries 29} (1998) 257}
  [\href{https://arxiv.org/abs/hep-ph/9701269}{{\ttfamily hep-ph/9701269}}].

\bibitem{Kitano:2004sv}
R.~Kitano and I.~Low, \emph{Dark matter from baryon asymmetry},
  \href{https://doi.org/10.1103/PhysRevD.71.023510}{\emph{Phys. Rev. D}
  {\bfseries 71} (2005) 023510}
  [\href{https://arxiv.org/abs/hep-ph/0411133}{{\ttfamily hep-ph/0411133}}].

\bibitem{Kaplan:2009ag}
D.E.~Kaplan, M.A.~Luty and K.M.~Zurek, \emph{Asymmetric dark matter},
  \href{https://doi.org/10.1103/PhysRevD.79.115016}{\emph{Phys. Rev. D}
  {\bfseries 79} (2009) 115016}
  [\href{https://arxiv.org/abs/0901.4117}{{\ttfamily 0901.4117}}].

\bibitem{Petraki:2013wwa}
K.~Petraki and R.R.~Volkas, \emph{Review of asymmetric dark matter},
  \href{https://doi.org/10.1142/S0217751X13300287}{\emph{International Journal
  of Modern Physics A} {\bfseries 28} (2013) 1330028}
  [\href{https://arxiv.org/abs/1305.4939}{{\ttfamily 1305.4939}}].

\bibitem{Zurek:2013wia}
K.M.~Zurek, \emph{Asymmetric dark matter: Theories, signatures, and
  constraints},
  \href{https://doi.org/https://doi.org/10.1016/j.physrep.2013.12.001}{\emph{Physics
  Reports} {\bfseries 537} (2014) 91}
  [\href{https://arxiv.org/abs/1308.0338}{{\ttfamily 1308.0338}}].

\bibitem{Iminniyaz:2011yp}
H.~Iminniyaz, M.~Drees and X.~Chen, \emph{Relic abundance of asymmetric dark
  matter}, \href{https://doi.org/10.1088/1475-7516/2011/07/003}{\emph{Journal
  of Cosmology and Astroparticle Physics} {\bfseries 2011} (2011) 003}
  [\href{https://arxiv.org/abs/1104.5548}{{\ttfamily 1104.5548}}].

\bibitem{Ghosh:2020lma}
A.~Ghosh, D.~Ghosh and S.~Mukhopadhyay, \emph{Asymmetric dark matter from
  semi-annihilation},
  \href{https://doi.org/10.1007/JHEP08(2020)149}{\emph{Journal of High Energy
  Physics} {\bfseries 2020} (2020) 149}
  [\href{https://arxiv.org/abs/2004.07705}{{\ttfamily 2004.07705}}].

\bibitem{Ghosh:2021ivn}
A.~Ghosh, D.~Ghosh and S.~Mukhopadhyay, \emph{{Revisiting the role of
  CP-conserving processes in cosmological particle\textendash{}antiparticle
  asymmetries}},
  \href{https://doi.org/10.1140/epjc/s10052-021-09848-5}{\emph{The European
  Physical Journal C} {\bfseries 81} (2021) 1038}
  [\href{https://arxiv.org/abs/2103.03650}{{\ttfamily 2103.03650}}].

\bibitem{Kolb:1979qa}
E.W.~Kolb and S.~Wolfram, \emph{Baryon number generation in the early
  universe},
  \href{https://doi.org/https://doi.org/10.1016/0550-3213(80)90167-4}{\emph{Nuclear
  Physics B} {\bfseries 172} (1980) 224}.

\bibitem{Dolgov:1979mz}
A.D.~Dolgov, \emph{{Baryon asymmetry of the universe and violation of the
  thermodynamics equilibrium. (in Russian)}}, {\emph{Pisma Zh. Eksp. Teor.
  Fiz.} {\bfseries 29} (1979) 254}.

\bibitem{Hook:2011tk}
A.~Hook, \emph{Unitarity constraints on asymmetric freeze-in},
  \href{https://doi.org/10.1103/PhysRevD.84.055003}{\emph{Phys. Rev. D}
  {\bfseries 84} (2011) 055003}
  [\href{https://arxiv.org/abs/1105.3728}{{\ttfamily 1105.3728}}].

\bibitem{Bhattacharya:2011sy}
A.~Bhattacharya, R.~Gandhi and S.~Mukhopadhyay, \emph{Revisiting the
  implications of {CPT} and unitarity for baryogenesis and leptogenesis},
  \href{https://doi.org/10.1103/PhysRevD.89.116014}{\emph{Phys. Rev. D}
  {\bfseries 89} (2014) 116014}
  [\href{https://arxiv.org/abs/1109.1832}{{\ttfamily 1109.1832}}].

\bibitem{Blazek:2021olf}
T.~Bla\v{z}ek and P.~Mat\'ak, \emph{{CP} asymmetries and higher-order unitarity
  relations}, \href{https://doi.org/10.1103/PhysRevD.103.L091302}{\emph{Phys.
  Rev. D} {\bfseries 103} (2021) L091302}
  [\href{https://arxiv.org/abs/2102.05914}{{\ttfamily 2102.05914}}].

\bibitem{Coster:1970jy}
J.~Coster and H.P.~Stapp, \emph{Physical‐region discontinuity equation},
  \href{https://doi.org/10.1063/1.1665443}{\emph{Journal of Mathematical
  Physics} {\bfseries 11} (1970) 2743}.

\bibitem{Bourjaily:2020wvq}
J.L.~Bourjaily, H.~Hannesdottir, A.J.~McLeod, M.D.~Schwartz and C.~Vergu,
  \emph{Sequential discontinuities of feynman integrals and the monodromy
  group}, \href{https://doi.org/10.1007/JHEP01(2021)205}{\emph{Journal of High
  Energy Physics} {\bfseries 2021} (2021) 205}
  [\href{https://arxiv.org/abs/2007.13747}{{\ttfamily 2007.13747}}].

\bibitem{Hannesdottir:2022bmo}
H.S.~Hannesdottir and S.~Mizera, \emph{{What is the i\ensuremath{\varepsilon}
  for the S-matrix?}}, SpringerBriefs in Physics, Springer (1, 2023),
  \href{https://doi.org/10.1007/978-3-031-18258-7}{10.1007/978-3-031-18258-7},
  [\href{https://arxiv.org/abs/2204.02988}{{\ttfamily 2204.02988}}].

\bibitem{Roulet:1997xa}
E.~Roulet, L.~Covi and F.~Vissani, \emph{On the {$CP$} asymmetries in majorana
  neutrino decays},
  \href{https://doi.org/https://doi.org/10.1016/S0370-2693(98)00135-X}{\emph{Physics
  Letters B} {\bfseries 424} (1998) 101}
  [\href{https://arxiv.org/abs/hep-ph/9712468}{{\ttfamily hep-ph/9712468}}].

\bibitem{Blazek:2021zoj}
T.~Bla{\v{z}}ek and P.~Mat{\'a}k, \emph{Cutting rules on a cylinder: a
  simplified diagrammatic approach to quantum kinetic theory},
  \href{https://doi.org/10.1140/epjc/s10052-021-09874-3}{\emph{The European
  Physical Journal C} {\bfseries 81} (2021) 1050}
  [\href{https://arxiv.org/abs/2104.06395}{{\ttfamily 2104.06395}}].

\bibitem{Blazek:2021gmw}
T.~Bla{\v{z}}ek and P.~Mat{\'a}k, \emph{Mass-derivative relations for
  leptogenesis},
  \href{https://doi.org/10.1140/epjc/s10052-022-10165-8}{\emph{The European
  Physical Journal C} {\bfseries 82} (2022) 214}
  [\href{https://arxiv.org/abs/2111.03419}{{\ttfamily 2111.03419}}].

\bibitem{Blazek:2022azr}
T.~Bla{\v{z}}ek, P.~Mat{\'a}k and V.~Zaujec, \emph{Mass-derivative relations
  and unitarity constraints for {$CP$} asymmetries at finite temperature},
  \href{https://doi.org/10.1088/1475-7516/2022/10/042}{\emph{J. Cosmol.
  Astropart. Phys.} {\bfseries 2022} (2022) 042}
  [\href{https://arxiv.org/abs/2209.03829}{{\ttfamily 2209.03829}}].

\bibitem{Botella:2004ks}
F.J.~Botella, M.~Nebot and O.~Vives, \emph{Invariant approach to
  flavour-dependent {CP}-violating phases in the {MSSM}},
  \href{https://doi.org/10.1088/1126-6708/2006/01/106}{\emph{Journal of High
  Energy Physics} {\bfseries 2006} (2006) 106}
  [\href{https://arxiv.org/abs/hep-ph/0407349}{{\ttfamily hep-ph/0407349}}].

\bibitem{Gondolo:1990dk}
P.~Gondolo and G.~Gelmini, \emph{Cosmic abundances of stable particles:
  Improved analysis},
  \href{https://doi.org/https://doi.org/10.1016/0550-3213(91)90438-4}{\emph{Nucl.
  Phys. B} {\bfseries 360} (1991) 145}.

\bibitem{Shtabovenko:2020gxv}
V.~Shtabovenko, R.~Mertig and F.~Orellana, \emph{{FeynCalc 9.3: New features
  and improvements}},
  \href{https://doi.org/https://doi.org/10.1016/j.cpc.2020.107478}{\emph{Computer
  Physics Communications} {\bfseries 256} (2020) 107478}
  [\href{https://arxiv.org/abs/2001.04407}{{\ttfamily 2001.04407}}].

\bibitem{Shtabovenko:2016sxi}
V.~Shtabovenko, R.~Mertig and F.~Orellana, \emph{{New developments in FeynCalc
  9.0}},
  \href{https://doi.org/https://doi.org/10.1016/j.cpc.2016.06.008}{\emph{Computer
  Physics Communications} {\bfseries 207} (2016) 432}
  [\href{https://arxiv.org/abs/1601.01167}{{\ttfamily 1601.01167}}].

\bibitem{Mertig:1990an}
R.~Mertig, M.~Böhm and A.~Denner, \emph{{Feyn Calc - Computer-algebraic
  calculation of Feynman amplitudes}},
  \href{https://doi.org/https://doi.org/10.1016/0010-4655(91)90130-D}{\emph{Computer
  Physics Communications} {\bfseries 64} (1991) 345}.

\end{thebibliography}\endgroup

\end{document}